\documentclass{jfm}

\usepackage{float}
\usepackage{graphicx}
\usepackage{natbib}
\usepackage{epstopdf}
\usepackage{breqn}
\usepackage{siunitx}
\usepackage{multirow}

\usepackage{color}
\ifCUPmtlplainloaded \else
  \checkfont{eurm10}
  \iffontfound
    \IfFileExists{upmath.sty}
      {\typeout{^^JFound AMS Euler Roman fonts on the system,
                   using the 'upmath' package.^^J}%
       \usepackage{upmath}}
      {\typeout{^^JFound AMS Euler Roman fonts on the system, but you
                   dont seem to have the}%
       \typeout{'upmath' package installed. JFM.cls can take advantage
                 of these fonts,^^Jif you use 'upmath' package.^^J}%
       \providecommand\upi{\pi}%
      }
  \else
    \providecommand\upi{\pi}%
  \fi
\fi

\ifCUPmtlplainloaded \else
  \checkfont{msam10}
  \iffontfound
    \IfFileExists{amssymb.sty}
      {\typeout{^^JFound AMS Symbol fonts on the system, using the
                'amssymb' package.^^J}%
       \usepackage{amssymb}%
         \let\leq=\leqslant
         
      }{}
  \fi
\fi

\ifCUPmtlplainloaded \else
  \IfFileExists{amsbsy.sty}
    {\typeout{^^JFound the 'amsbsy' package on the system, using it.^^J}%
     \usepackage{amsbsy}}
    {}
\fi

\newcommand\mum{\nobreak\mbox{$\mu$m}}

\newcommand\Pran{\mbox{\textit{Pr}}} 
\newcommand\Ma{\textit{Ma}}					 
\newcommand\Biot{\textit{Bi}}				 
\newcommand\Ca{\textit{Ca}}					 
\newcommand\Bo{\textit{Bo}}					 
\newcommand\Dy{\textit{Dy}}					 
\newcommand\Ga{\textit{Ga}}					 
\newcommand\Ra{\textit{Ra}}					 

\newsavebox{\astrutbox}
\sbox{\astrutbox}{\rule[-5pt]{0pt}{20pt}}

\title[Conjugated liquid layers driven by the B\'enard-Marangoni instability]{Conjugated liquid layers driven by the short-wavelength B\'enard-Marangoni instability: experiment and numerical simulation}

\author[I. Nejati , M. Dietzel and S. Hardt]%
{Iman Nejati,\ns 
 Mathias Dietzel 
\thanks{Email address for correspondence: dietzel@csi.tu-darmstadt.de}\ns
\break
and Steffen Hardt }

\affiliation{Institute for Nano- and Microfluidics, Center of Smart Interfaces, TU Darmstadt,
Alarich-Weiss-Strasse 10, 64287, Darmstadt, Germany\\[\affilskip]}

\pubyear{2010}
\volume{650}
\pagerange{119--126}
\date{?; revised ?; accepted ?. - To be entered by editorial office}
\begin{document}

\maketitle

\begin{abstract}
The coupled dynamics of two conjugated liquid layers of disparate thicknesses, which coat a solid substrate and are subjected to a transverse temperature gradient, is investigated. The upper liquid layer evolves under the short-wavelength (SW) B\'enard-Marangoni (BM) instability, whereas the lower, much thinner film undergoes a shear-driven long-wavelength deformation. Although the lubricating film should reduce the viscous stresses acting on the up to one hundred times thicker upper layer by only $10\%$, it is found that the critical Marangoni number of marginal stability may be as low as if a stress-free boundary condition were applied at the bottom of the upper layer, i.e. much lower than the classical value of $79.6$ known for a single film. Furthermore, it is experimentally verified that the deformation of the liquid-liquid interface, albeit small, has a non-negligible effect on the temperature distribution along the liquid-gas interface of the upper layer. This stabilizes the hexagonal pattern symmetry towards external disturbances and indicates a two-way coupling of the different layers. The experiments also demonstrate how convection patterns formed in a liquid film can be used to pattern a second conjugated film. The experimental findings are verified by a numerical model of the coupled layers.
\end{abstract}

\begin{keywords}
Coupled liquid layers, thin film flow, short-wavelength instability, long-wavelength deformation, B\'enard-Marangoni instability \\
\indent{\rule{12.7cm}{0.4pt}} \\ \\
This is the post-print authors' version of the manuscript, which was published in the Journal of Fluid Mechanics.
\copyright \ Cambridge University Press 2015. doi:10.1017/jfm.2015.544
\end{keywords}

\section{Introduction}\label{sec:Introduction}
The principles of self-organization, referring to the evolution of a non-equilibrium system to an ordered state in the absence of external agents, are omnipresent in nature and technological entities \citep{nicolis1977self}. There are many manifestations of self-organization in hydrodynamic and diffusion-driven instabilities \citep{bestehorn06}. One of the most well-known examples is pattern formation driven by thermal convection. A century after \citet{Ben1900} discovered cellular convective structures in liquid layers subjected to a temperature gradient, thermal convection in fluid layers is still a subject of intense research. Next to the practical relevance for a variety of applications, there is a fundamental interest in the evolution dynamics and (self-organized) pattern formation of liquid interfaces. Within this scope, surface tension-driven instabilities become dominant for layer thickness below \textit{O}(\SI{1}{mm}). Sparked by the seminal work of \citet{Pearson58}, numerous studies have been devoted to the B\'enard-Marangoni (BM) instability in a liquid layer, which is driven by a transverse temperature gradient in conjunction with a temperature-dependent surface tension (thermocapillarity) \citep[see][]{Kosch86,Colinet05,Davis87}. In general, thermocapillarity can cause two fundamentally different instability modes: For thicker films, the short-wavelength (SW) mode generates cellular convection patterns, whose horizontal periodicity is of similar magnitude as the layer thickness. By contrast to the SW-mode, which can develop without any interfacial deformation, the long-wavelength (LW) mode always goes along with interfacial deformations and is dominant in thinner or more viscous films. The deformation amplitude is typically much smaller than the pattern periodicity but may reach a similar magnitude as the film thickness. The onset of instability, pattern symmetry and dynamics of a liquid layer undergoing the SW-BM instability have been analyzed both experimentally and theoretically by a number of researchers. For instance, \citet{Schatz95} used shadowgraphy to visualize the pattern and analyze hysteresis effects during heating and cooling cycles. \citet{Rahal07} have studied the pattern dynamics, the (weak) free surface deformation as well as the interfacial temperature field of SW-BM convection in a circular container, using interferometry and infrared thermography. The SW-BM convection is relevant for a number of practical problems such as crystal growth \citep{Hurle81}, welding of steels \citep{Mills90} and convection under microgravity \citep{Op11}. With respect to the LW-BM instability, in their experimental as well as theoretical study of \textit{O}(\SI{100}{\mum}) thick oil films, \citet{Vanhook97} have reported that, depending on the thicknesses and thermal conductivities of the liquid and the gas layer, either localized depressions or elevations will form at the interface. \citet{McLeod11} showed that the LW-BM instability can occur in \textit{O}(\SI{100}{nm}) polymer film melts, which was also reported for less than \SI{20}{nm} thick metal film melts by \citet{Trice08}. In addition, \citet{Burgess01} demonstrated that thermocapillary stresses may suppress the LW-Rayleigh-Taylor (RT) instability.
  
While a large number of studies were conducted investigating the behavior and characteristics of fluid layers undergoing an instability driven by a single mechanism, only a few studies were carried out specifically with the intention to address the interaction between different forms of instabilities. \citet{Golovin94} have studied the interplay between SW- and LW-Marangoni convection in a film subjected to a transverse concentration gradient. They showed that in the case when both modes are individually unstable, SW-convection can stabilize the LW-deformation and - under certain conditions - changes its type to oscillatory instead of monotonic. Furthermore, they indicated that the LW-mode can trigger the SW-convection instability by surface deformations. At present, work on systems with multiple (immiscible) liquid layers is mainly concerned with configurations where each layer has a similar thickness so that the same type of instability (SW or LW) is dominant in each layer. To the best of our knowledge, the earliest works in this field have been conducted for thick stratified liquid layers heated from below \citep[see][]{Andereck98}. Motivated by the thermal convection present in the Earth's mantle, the liquid system investigated was set up in such a fashion that buoyancy alongside thermocapillarity at the liquid-liquid interface drives the SW-convection instability. In their study on the coupling between two superimposed liquid layers both undergoing SW-Rayleigh-B\'enard (RB) instabilities, \citet{Rasenat89} reported a qualitatively different behavior of the conjugated system compared to the behavior of the individual single films. They identified three different coupling modes between the layers, namely: viscous, thermal and oscillatory. \citet{Colinet94} showed that the viscous mode may appear for thickness ratios (upper to lower layer) smaller than one, whereas thermal coupling dominates for thickness ratios larger than one, depending on the physical properties of the liquids. They pointed out that the competition between these two mechanisms leads to time-periodic oscillatory motion if both layers are of similar thickness. These coupling modes have been analyzed and observed experimentally, in agreement with the theoretical predictions \citep{Kang2003,Prakash97}. By sufficiently heating from above, \citet{Welander64} has shown that in a two-layer system the liquid layers undergo a type of convective instability, which is called ''anti-convection'', if the thermal expansion coefficient and the conductivity of one layer is much larger compared to the other one \citep[see][]{Merkt12,H14}.

Researchers have also studied systems where both layers are rather thin and dominated by LW-modes. For instance, in their study of the thermal coupling between two thin liquid layers undergoing LW-RT as well as LW-BM instabilities, \citet{Vecsei14} have reported substantially different stability properties in the combined system compared to the uncoupled case. \citet{Merkt2005} have reported that thermocapillarity at the interface between two thin liquid layers undergoing a LW-instability may have a stabilizing or destabilizing effect, depending on the material properties. They emphasized that both heating from below or from above may destabilize the system. Furthermore, they have shown that a RT-unstable two-layer system can be stabilized by heating from below.

In all of these studies it has been shown that the interplay between liquid layers subjected to temperature or concentration gradient may give rise to some new characteristics compared to systems with only a single liquid layer. However, so far the scientific community has apparently neglected systems of multiple liquid layers within a temperature gradient where the layer thicknesses substantially differ from each other, so that SW-modes dominate in one and LW-modes in the other. Based on this background, in this work the interaction between an upper liquid layer with a conjugated but much thinner lower liquid film, both exposed to a transverse temperature gradient, is experimentally studied. The upper liquid layer is seen to undergo a SW-BM instability, which induces LW-deformations of the lower liquid film through viscous stresses at the liquid-liquid interface. It is shown that not only the reduced friction at the bottom of the upper layer affects its stability behavior but, counter-intuitively, also the small deformations at the liquid-liquid interface. Such an effect has never been described before. The experimental setup and procedure as well as the experimental results are described in $\S 2$. A two-dimensional ($2$-D) mathematical model has been developed to analyze the behavior of the coupled system, which is described in $\S 3$, followed by a comparison of the experimental and the numerical results. 

\section{Experiment}
\subsection{Experimental setup}
\begin{figure}
 \centering
  \includegraphics{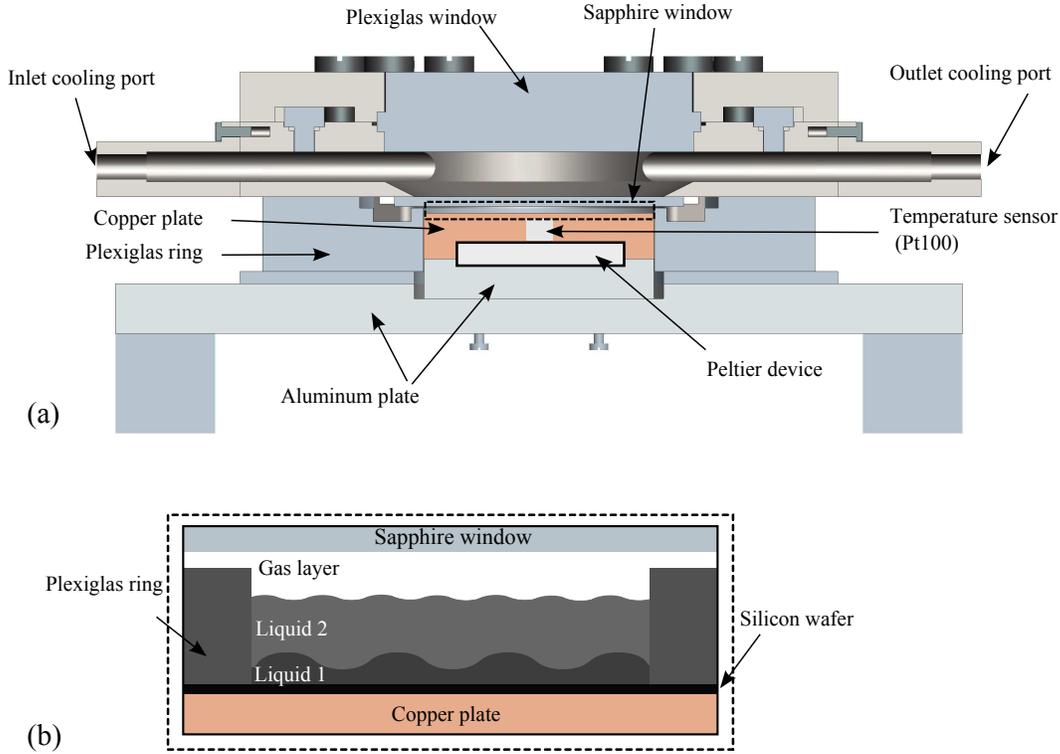}
   \caption{(a) Cross-section of the experimental setup (drawn to scale). (b) Those parts of the experimental setup located inside the rectangle indicated by a dashed line in (a) (magnified view). A thin silicon wafer placed on top of the copper plate is used as a light-reflecting surface. The Plexiglas ring is placed onto the silicon wafer whose inner area is smaller than the area of the Peltier device. This ensures a homogeneous temperature distribution beneath the liquid films. The contact area between the Plexiglas ring and the silicon wafer has been sealed.}
\label{fig:ExSet}
\end{figure}
As schematically illustrated in figure \ref{fig:ExSet}, the experimental cell contains two stacked liquid layers, which are exposed to a transverse temperature gradient. Several factors have to be taken into account to select proper working fluids, including immiscibility, vapor pressure under ambient conditions as well as optical and wetting properties. In order to analyze the interaction between SW- and LW-modes of interfacial instabilities, the thickness of the lower layer (\SI{5}{\mum}) is selected to be much smaller than that of the upper one (a few hundred micrometers). For the lower layer, silicone oil (\SI{3}{$\textrm{cSt}$}, Elbesil) has been chosen as it completely wets most surface materials. This implies that very small film thicknesses can be achieved without film rupture being observed. Perfluorinated hydrocarbon (FC-70, Sigma-Aldrich) has been chosen as the upper layer, since it is immiscible with silicone oil and has a low surface tension with the oil. The thermophysical properties of the liquids - as provided by the manufacturers - are listed in table 1. A layer of air (thickness ranges from approximately \SI{1200}{\mum} to \SI{1700}{\mum}) separates the liquid layers from the upper substrate.
A silicon wafer (diameter \SI{5}{$\textrm{cm}$}) is used as the lower substrate and placed on a \SI{10}{$\textrm{mm}$} thick copper plate whose lower surface is heated by a \SI{32}{$\textrm{W}$} Peltier device. Since FC-70 has a higher density than the silicone oil (see table 1), it must be deposited very gently onto the lower film to prevent film rupture. As will be detailed later, in spite of the density inversion in the layered system, the liquid-liquid interface remains flat as long as the oil film is sufficiently thin.
The conjugated liquid layers are kept inside a ring of Plexiglas with an inner diameter of \SI{38}{$\textrm{mm}$} and a height of \SI{1}{$\textrm{mm}$}. The ring is placed on top of the lower substrate and is sealed with a silicone-based sealing compound (Dirko HT). The temperature of the copper plate is measured with a platinum resistance temperature detector (RTD, Pt100) and is accurate to within $\pm 0.15\:^\circ\textrm{C}$. The required power for the Peltier device to maintain the temperature of the copper plate at a specified value is supplied by a Newport temperature controller (model 3040). In order to minimize the heat loss to the environment, the copper plate is surrounded by a ring of Plexiglas (thermal conductivity of $0.17\:\textrm{W}\textrm{m}^{-1}\textrm{K}^{-1}$) with an outer diameter of \SI{12}{$\textrm{cm}$} (see figure \ref{fig:ExSet}).
\begin{figure}
  \centerline{\includegraphics{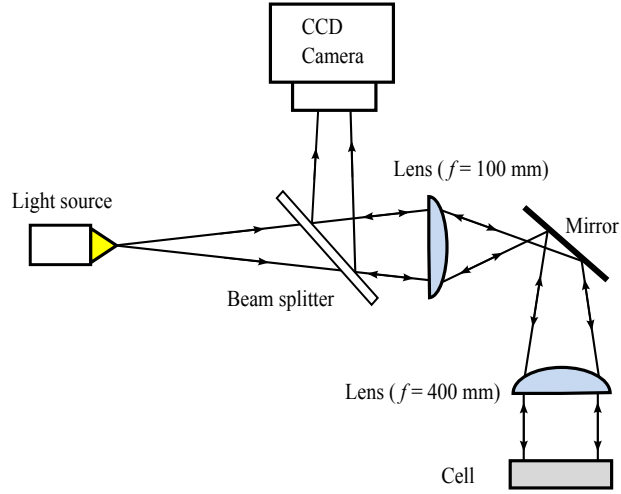}}
  \caption{Sketch of the employed shadowgraphy system}
\label{fig:Shadowgraph}
\end{figure}
The cooling system is designed to allow optical access to the liquid layers from the top. For this purpose, a \SI{2}{$\textrm{mm}$}-thick sapphire window (diameter \SI{58}{$\textrm{mm}$}) is employed as an upper substrate and cooled by the circulating flow of a cooling fluid (type Thermal G, Julabo). The latter is temperature-controlled by a thermostat (VWR 1190P, temperature stability $\pm 0.01\:^\circ\textrm{C}$) and passes in between the sapphire and another Plexiglas window (\SI{15}{$\textrm{mm}$} thick). The temperature distribution along the sapphire glass can be assumed to be quasi-homogeneous since the sapphire window exhibits a high thermal conductivity of \SI{46}{$\textrm{W}\textrm{m}^{-1}\textrm{K}^{-1}$} and is continuously and uniformly cooled by a high flow rate of the coolant. Furthermore, the thick Plexiglas window minimizes heat input to the cooling section from the environment. A temperature sensor (Pt100) is utilized to measure the temperature of the sapphire window.

The total gap between the sapphire window and the silicon wafer can be adjusted using four screws installed on the \SI{10}{$\textrm{mm}$} aluminum plate beneath the experimental cell. In this paper the total gap width is constant and equals $2230\:\mu\textrm{m}\pm10\:\mu\textrm{m}$. The gap uniformity (parallelism between the lower and the upper substrate) lies within $2\%$. The total thickness of the liquid layers is measured using a stylus attached to a micrometer caliper, which has an accuracy of $\pm$\SI{5}{\mum}. The position of the liquid-gas interface is determined by the instance when the liquid begins to wet the tip of the stylus, while the bottom of the lower liquid film is determined by the contact of the stylus with the silicon wafer (signaled through electric short-circuiting). This measurement is carried out without transverse temperature gradient ($\Delta T=0$) since thermocapillarity can affect the wetting of the stylus by the liquid. Here $\Delta T=T_h-T_c$, where $T_c$ and $T_h$ are the temperatures of the upper and the lower substrate, respectively. The average value of the lower liquid film thickness is approximately $h_{01}\approx$ \SI{5.0}{\mum} $\pm$ \SI{0.1}{\mum}, which is deduced from the division of the initially deposited volume by the surface area of the lower substrate.
The depth of the upper liquid layer is obtained by subtraction of the lower film thickness from the total thickness of the liquid layers and ranges from $h_{02}\approx \SI{400}{\mum}$ to \SI{1000}{\mum}.

The characteristic time of thermal diffusion in vertical direction in the upper layer can be approximated by $\tau_v=h_{02}^2/\alpha_2$, where $\alpha_2$ is the thermal diffusivity of the upper layer. For FC-70 and the largest value of $h_{02}$ used in this study, one finds $\tau_v\approx 32\:\textrm{s}$. Similarly, the thermal diffusion time in horizontal direction is $\tau_h=4\Gamma^2\tau_v$, where $\Gamma$ represents the ratio of the inner radius of the Plexiglas ring and the height of the upper liquid layer and has a maximum value of $47.5\pm0.05$. In order to perform a quasi-stationary analysis the rate by which the temperature difference is increased has been chosen much less than $(1/\tau_v)$, while reducing this value below $(1/\tau_h)$ was found to be impractical. Hence, the overall temperature gradient is ramped up slowly with a slope of $0.05\:\textrm{K min}^{-1}$, so that imperfect patterns - possibly appearing in the experiments due to transient effects - are minimized.

The visualization of the film configuration is achieved using standard shadowgraphy (figure \ref{fig:Shadowgraph}), which is based on the change of the refractive index with temperature \citep{Merzkirch87,Eckert98,Degen98}. Quasi-parallel light is obtained from a point-like light source by using two optical lenses (with diameters of \SI{25}{\textrm{mm}} and \SI{50.8}{\textrm{mm}} and focal distances of \SI{100}{\textrm{mm}} and \SI{400}{\textrm{mm}}, respectively) and projection onto the liquid surface parallel to the surface normal. The illumination light is reflected from the silicon wafer and diverted by a beam splitter. This produces a two dimensional gray value distribution which is recorded by a CCD camera (Nikon DS-Qi1Mc, resolution $1280$x$1024$).

\begin{table}
  \begin{center}
\def~{\hphantom{0}}
  \begin{tabular}{lcll}
      Parameter  					              & Symbol     &FC-70 ($25\:^\circ\textrm{C}$) 										         & Silicone oil ($25^\circ\textrm{C}$) \\[3pt]
       Density   					              & $\rho$     &1940$\:\textrm{kg m}^{-3}$ 						  	                 &925$\:\textrm{kg m}^{-3}$\\
       Surface tension (with air)       & \multirow{2}{*}{$\gamma$}   &0.018$\:\textrm{N m}^{-1}$						  	                 &0.019$\:\textrm{N m}^{-1}$\\
       Interfacial tension (with silicone oil)&          &0.005$\:\textrm{N m}^{-1}$						  	                 &\multicolumn{1}{c}{-}\\
       Marangoni coefficient (with air)	&\multirow{2}{*}{$\gamma_T$}  &-6.7$\times10^{-5}\:\textrm{N m}^{-1}\textrm{K}^{-1}$      &-6.2$\times10^{-5}\:\textrm{N m}^{-1}\textrm{K}^{-1}$\\
       Marangoni coefficient (with silicone oil) &  &-3.2$\times10^{-5}\:\textrm{N m}^{-1}\textrm{K}^{-1}$       &\multicolumn{1}{c}{-}\\
       Thermal conductivity             & $\kappa$   &0.07$\:\textrm{W m}^{-1}\textrm{K}^{-1}$  			           &0.125$\:\textrm{W m}^{-1}\textrm{K}^{-1}$\\
       Thermal diffusivity              & $\alpha$   &3.28$\times10^{-8}\:\textrm{m}^{2}\textrm{s}^{-1}$         &8.60$\times10^{-8}\:\textrm{m}^{2}\textrm{s}^{-1}$\\
       Thermal expansion coefficient    & $\beta$    &1.00$\times10^{-3}\:\textrm{K}^{-1}$                       &1.10$\times10^{-3}\:\textrm{K}^{-1}$\\
       Kinematic viscosity              & $\nu$      &12.0$\times10^{-6}\:\textrm{m}^{2}\textrm{s}^{-1}$         &3.00$\times10^{-6}\:\textrm{m}^{2}\textrm{s}^{-1}$\\
       
  \end{tabular}
  \caption{Physical properties of FC-70 and silicone oil (as obtained from the manufacturers)}
  \label{tab:pp}
  \end{center}
\end{table}

\subsection{Experimental results}
\label{exp_result}

\begin{figure}
  \centerline{\includegraphics{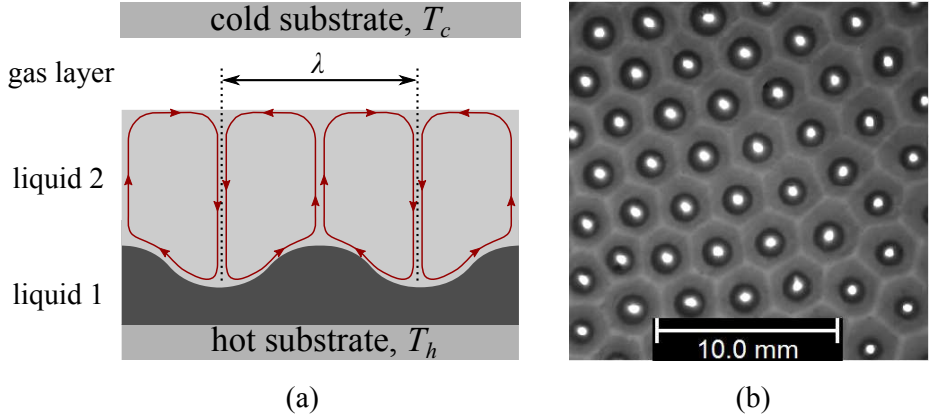}}
  \caption{(a) Schematic of the coupled configuration of two conjugated liquid layers (side view). The streamlines in the upper liquid film are sketched as solid lines, while the edges of the convection cells are illustrated as dotted lines. (b) Snapshot of a typical experimental result (top view). The bright lines are edges of the convection cells emerging in the upper layer, whereas the bright spots at the cell centers are bulges of the lower liquid film. Here, $h_{02}=960\:\mu\textrm{m}$, and the temperature difference across the upper layer is $\Delta T_{2}=4.4\:\textrm{K}$ (corresponds to $\Ma=390$).}
\label{fig:coupl}
\end{figure}

In all performed experiments, it has been observed that the films in the conjugated system evolve according to a SW-interfacial instability coupled to a LW-interfacial deformation, where the horizontal extent of the deformation is much larger than the thickness of the lower film. This is illustrated in figure \ref{fig:coupl}(a). The dynamic behavior of the coupled system under study is mainly determined by gravitational and thermocapillary effects as well as by thermal diffusion in each individual layer. Accordingly, the relevant time scales are $\tau^2_{g,i}=h_i/g$, $\tau^2_{\textrm{ther},i}=\rho_i h_i^3/(\gamma_{T,i} \Delta T_i)$ and $\tau^2_{\textrm{diff},i}=h_i^4/(\nu_i\alpha_i)$ \citep{Vanhook97}. Here, $\nu_i$, $\mu_i$ and $\rho_i$ denote the kinematic viscosity, dynamic viscosity and density of liquid layer $i$, respectively. The subscript $i=1,2$ refers to the lower ($i=1$) and the upper layer ($i=2$), respectively. The gravitational acceleration is expressed by $g$, while $\gamma_{T,i}=|d\gamma_i/dT|$, where $\gamma_i$ is the interfacial tension of interface $i$. Here, $i=1$ refers to the liquid-liquid interface, while $i=2$ refers to the liquid-gas interface. The temperature difference across liquid layer $i$ is denoted by $\Delta T_i$. The significance of buoyancy in the upper liquid layer can be estimated with the Rayleigh number $\Ra=\beta_2 g h_{02}^3\Delta T_2/(\nu_2\alpha_2)$, i.e. with the ratio of the thermal diffusion $\tau^2_{\textrm{diff},2}$ and the buoyancy time scale $\tau^2_{\textrm{buo},2}=h_{02}/(\beta_2 g \Delta T_2)$. Here, $\beta_2$ denotes the thermal expansion coefficient of the upper liquid layer. Since $h_{02}$ is sufficiently small, $\Ra$ is much smaller than the critical value at marginal stability of RB-convection: the maximum value of the $\Ra$ is approximately $45$, which is much smaller than the critical value for a stress-free ($657.5$) or a non-slipping ($1700$) lower boundary \citep{chandrasekhar1970hydrodynamic}. The ratio of the thermal diffusion and the gravitational time scale in the upper liquid layer (defined as the Galileo number $\Ga=gh_{02}^3/(\nu_2\alpha_2)$) is sufficiently large to prevent the growth of deformational perturbations at the liquid-gas interface: Large $\Ga$-numbers imply that the deformational (LW) mode of the BM-instability can only occur if the critical threshold for the SW-mode is exceeded as well. However, SW-instabilities imply that the liquid pressure is not uniform across the layer thickness, preventing the growth of deformational perturbations due to stabilization by means of the hydrostatic pressure. Consequently, the liquid-gas interface virtually remains flat. If the Marangoni number in the upper liquid layer (identical to the ratio of the thermal diffusion and thermocapillary time scale) $\Ma=-\gamma_{T,2} \Delta T_{2} h_{02} /(\mu_2\alpha_2)$ exceeds a critical threshold, thermocapillarity is strong enough to overcome stabilization by thermal diffusion and viscous dissipation, so that the SW-BM instability is initiated in the upper layer \citep{Schatz2001}. Here, $\gamma_{T,2}$ denotes the change of the interfacial tension with temperature at the liquid-gas interface (Marangoni coefficient).

\begin{figure}
  \centerline{\includegraphics{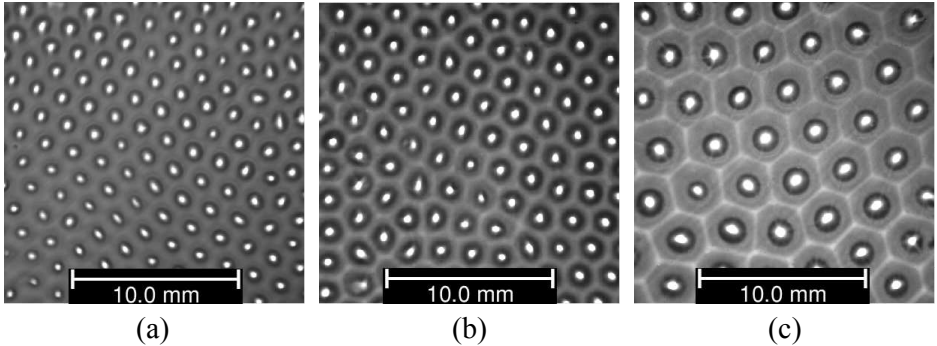}}
  \caption{Wavelength of the pattern in the lower film changing with variation of the upper film thickness. The thickness of the upper film as well as the characteristic temperature difference across it are (a) $h_{02}=510\:\mu\textrm{m}$, $\Delta T_{2}=2.5\:^\circ\textrm{C}$ ($\Ma=112$, $\lambda\approx1.6\:\textrm{mm}$) (b) $h_{02}=610\:\mu\textrm{m}$, $\Delta T_{2}=3\:\textrm{K}$ ($\Ma=163$, $\lambda\approx1.8\:\textrm{mm}$) and (c) $h_{02}=970\:\mu\textrm{m}$, $\Delta T_{2}=4\:\textrm{K}$ ($\Ma=341$, $\lambda\approx2.9\:\textrm{mm}$). In all cases, the thickness of the lower film is $5\:\mu\textrm{m}$.}
\label{fig:WL}
\end{figure}

The SW-convection cells in the upper liquid layer are classified as l-type instabilities, as they are formed by warm liquid rising upwards in the center of the cell and cooler fluid flowing downwards along the rims \citep{Thess95}. This is experimentally verified by adding small, density-matched tracer particles (S-HGS-10, diameter $10\:\mu\textrm{m}$, Dantec Dynamics GmbH) to the upper liquid layer. By means of viscous stresses, the convection shears and deforms the liquid-liquid interface towards the center of the cell. The deformations are seen to follow the hexagonal symmetry of the velocity field in the convection cells of the upper layer. A typical example of the shear-driven deformation of the liquid-liquid interface is shown in figure \ref{fig:coupl}(b). While the liquid-liquid interface is elevated in the center of the convection cells emerging in the upper liquid layer, the thickness of the lower liquid film decreases considerably close to the rims. Here, molecular interactions between the liquid-liquid and the solid-liquid interface in form of disjoining pressure effects become relevant and prevent film rupture. This was verified by a computational model described in a later section. The characteristic wavelength of the pattern in the upper layer is mainly a function of the liquid thickness $h_{02}$, and increasing the latter leads to the formation of wider convection cells. For any horizontal extent of the specific convection cells observed in the experiments, the viscous stresses in the upper liquid layer lead to a LW-deformation of the lower film. This can be seen in figure \ref{fig:WL}.

In the lower film the thermal diffusion time scale is sufficiently small, so that convection-induced temperature fluctuations are quickly damped. As a consequence, the corresponding Marangoni number is far below the critical value of SW-BM convection and the corresponding instability cannot develop in the lower film. By contrast, the growth rate of deformational perturbations depends on the ratio of the gravitational and thermocapillary time scale, which is defined as the inverse dynamic Bond number $\Dy_1\equiv\Ma_1/\Ga_1=-\gamma_{T,1}\Delta T_1/(\rho_1 g h_{01}^2)$. Here, $\gamma_{T,1}$ represents the change of the interfacial tension at the liquid-liquid interface with temperature. When the inverse dynamic Bond number is larger than $\Dy_c=2/3$, thermocapillarity is large enough to overcome gravitational stabilization and the liquid film undergoes the LW-thermocapillary instability \citep{Smith66,Vanhook97}. Despite that the lower film has a small thermal resistance and the temperature difference across this layer is comparably small, the inverse dynamic Bond number can nevertheless exceed $\Dy_c$ if the thickness of the lower film is sufficiently small. In this study the inverse dynamic Bond number of the lower film takes a maximum value of $\Dy_{1,max}=1.5$. However, as detailed in appendix \ref{appA}, the deformation and growth rate of the liquid-liquid interface due to the LW-thermocapillary instability are much smaller than those caused by the emerging shear-driven flow in the time frame of interest. 

In this context, it has to be noted that the stacked system is density-inverted, as the upper liquid has a higher density than the lower one (see table \ref{tab:pp}). Thus, in the absence of temperature gradients, the lower liquid film can undergo a LW-RT instability \citep{chandrasekhar1970hydrodynamic}. For a sufficiently thin lower film this is effectively damped or even suppressed altogether by disjoining pressure effects. A linear stability analysis (not shown herein for brevity) indicates that for film thicknesses below $h_{01C}=850\:\textrm{nm}$, the LW-RT instability is suppressed for the selected liquid-solid combination. As the thickness of the lower film exceeds $h_{01C}$, in the absence of a temperature gradient some undulations caused by the LW-RT instability can be observed at the liquid-liquid interface. However, the growth rate of these deformations strongly depends on the initial thickness of the lower film. Since this thickness is small in the system under study, the growth rate of the deformations is small as well; even after several hours no specific change at the liquid-liquid interface can be observed. This finding is confirmed by corresponding simulations discussed in appendix \ref{appA}, which are based on the lubrication approximation detailed in section 3. From the simulations it is also found that the amplitude of these deformations is much smaller than that of the shear-driven deformations. Since the coefficients of thermal expansion of FC-70 and silicone oil are almost identical ($\beta\approx 0.001\:\textrm{K}^{-1}$), the presence of a temperature gradient does not make it more likely for the LW-RT instability to have a non-negligible contribution to the deformation of the lower film interface. 
 
\begin{figure}
  \centerline{\includegraphics{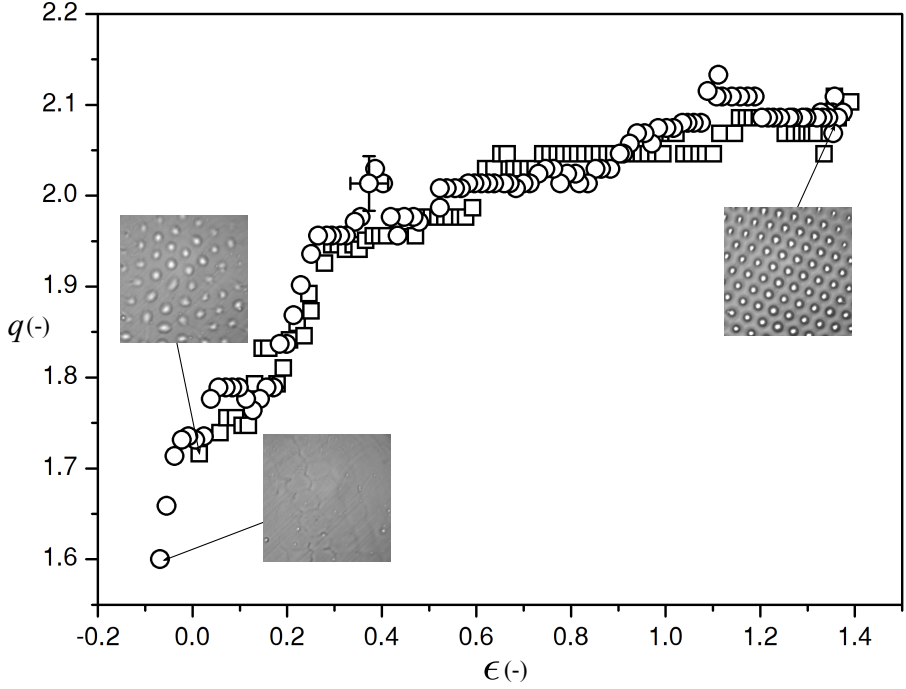}}
  \caption{The dimensionless wave number $q=2\pi h_{02}/\lambda$ as a function of $\epsilon=\Delta T_{2}/\Delta T_{2C} -1$ after the onset of instability in the upper liquid layer. Here $\lambda$ denotes the wavelength of the pattern. The initial thickness of the upper layer is $640\:\mu\textrm{m}$. Open squares correspond to data obtained by increasing $\epsilon$, while open circles correspond to data obtained by decreasing $\epsilon$, both in a quasi-stationary manner. Photographs of the corresponding film structure when viewed from above are displayed for three ($q$,$\epsilon$)-pairs (insets).}
\label{fig:Hyst}
\end{figure}

Equivalent to the behavior of a single liquid layer undergoing a SW-BM instability, the structure of the pattern emerging at the liquid-liquid interface depends on the control parameter $\epsilon=\Delta T_{2}/\Delta T_{2C} -1$, where $\Delta T_{2C}$ corresponds to the temperature difference at the onset of the BM-instability in the upper liquid layer. The dimensionless wave number of the pattern developing at the liquid-liquid interface of wavelength $\lambda$, denoted by $q=2\upi h_{02}/\lambda$, as a function of $\epsilon$ is shown in figure \ref{fig:Hyst}. In order to determine $q$ from the shadowgraphy images recorded, the latter have been transformed from position space to Fourier space. To this end, the gray-scale distribution of the digitized image was Fourier-transformed, giving the power spectral density $P(q,\phi)$, where $q$ and $\phi$ are polar coordinates in wave number space. Before Fourier-transforming, the intensity of the gray-scale images was rescaled to lie between $0$ and $255$. The raw value of $P(q,\phi)$ was corrected by a low-pass filter to cut off higher harmonic ringing, whereas (low-intensity) LW-noise has been removed using a high-pass filter. With $\epsilon$ in the upper liquid layer increasing to positive values, convection cells are formed, which deform the liquid-liquid interface. As the temperature gradient is slowly increased, the wavelength of the convection cells in the upper liquid layer decreases gradually, which manifests itself also in the deformational pattern at the liquid-liquid interface. The maximum value of $\epsilon$ considered in the experiments is selected in such a fashion that the hexagonal pattern is preserved, i.e. no bifurcation to a different pattern (squares, rolls) takes place. By decreasing the temperature gradient in a quasi-stationary manner, the shear stress at the liquid-liquid interface decreases gradually. Subsequently, the capillary pressure at the liquid-liquid interface flattens the deformations. At a specific value of $\epsilon$, the pattern symmetry and mean wave number found by reducing the temperature gradient is practically identical to the one emerging by increasing the temperature gradient. This is an indication for the fact that the temperature gradient was increased/decreased sufficiently slowly. However, after a complete cycle of $\epsilon$ the deformations of the liquid-liquid interface do not completely disappear, even for slightly negative values of the control parameter. In the system under study, this hysteresis is due to the presence of the RT instability, which becomes dominant before and after the temperature cycling, i.e. in the absence of the shear stresses exerted by the upper liquid layer. As discussed earlier, deformations of the liquid-liquid interface caused by this instability evolve much slower than those driven by the shear stress generated in the upper liquid layer. A mismatch of the dominant mean wave number before and after a temperature cycle cannot only be found in systems with conjugated films. For instance, it is known that also for a single liquid layer the optical pattern caused by the convection cells does not completely vanish after finishing a temperature cycle \citep{Schatz2001}. This is commonly attributed to hysteresis effects due to nonlinear symmetry breaking \citep{Cross93}, without elucidating this issue in great detail.

\begin{figure}
  \centerline{\includegraphics{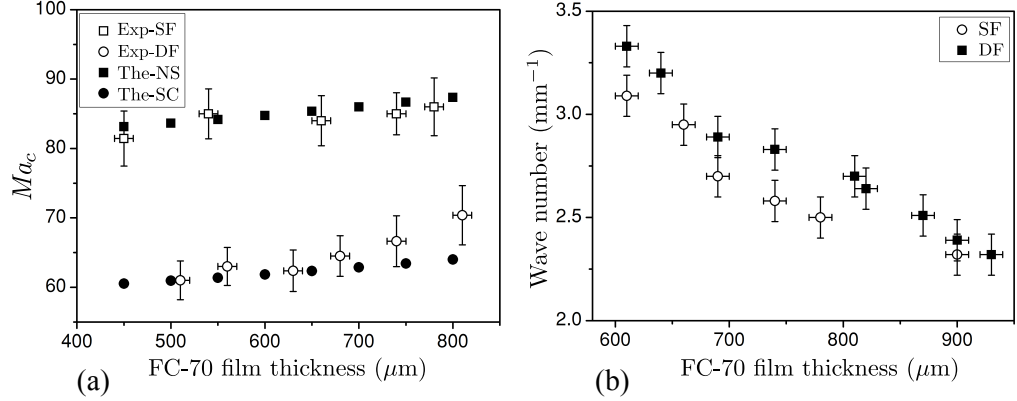}}
  \caption{(a) The onset of the SW-instability in a single film (SF) compared to the one in the stacked configuration (DF). For comparison with the experimental results (Exp), in the theoretical analysis (The), a no-slip condition (NS) or a free-slip condition (SC) is applied at the bottom of the thicker liquid layer. (b) Dominant wave number of the pattern as a function of the upper liquid thickness for a single film and the stacked configuration. All data shown in (b) were obtained for $\Delta T_{2}=3\:\textrm{K}$, which is above the point of marginal stability. The error bars are calculated based on an error propagation method incorporating uncertainties in the temperature as well as in the film thickness measurements, whereas the uncertainties of the values used for the relevant physical properties have been neglected.}
\label{fig:Ons_WL}
\end{figure}

The first theoretical investigation of the onset of thermocapillary convection in the absence of buoyancy forces for a liquid film coating a heated wall was conducted by \citet{Pearson58}. He found that the instability occurs when the Marangoni number exceeds the critical value of $\Ma_c\approx79.6$. This value corresponds to the case of a non-deformable liquid-gas interface and a no-slip as well as a no-penetration boundary condition applied at the bottom wall. Moreover, the critical Marangoni number is a function of the Biot number $\Biot=ah/\kappa$, where $a$ represents the heat transfer coefficient at the liquid-gas interface and $\kappa$ denotes the thermal conductivity of the liquid. The value of $\Ma_c\approx79.6$ corresponds to an adiabatic interface ($\Biot=0$). For the purpose of validation of the experimental setup, in figure \ref{fig:Ons_WL}(a) the critical Marangoni numbers experimentally obtained for a single layer of FC-70 (without the presence of the thin oil film underneath) are compared with the theoretical results of \citet{Pearson58} for different Biot numbers. It can be seen that the experimental results agree quite well with the theoretical predictions. Given this agreement between our experimental findings and the well-known theoretical results, the experimental setup used in this study should be suitable for analyzing the effects introduced by the thin oil film in a quantitative manner. In the case of the two conjugated liquid layers heated from below, at the liquid-liquid interface the tangential viscous stress from the lower liquid film is balanced by the sum of the tangential viscous stress of the upper film and the thermocapillary stress. The latter is small in comparison to the viscous stresses: Given that the thermal resistance of the thin oil layer is very small, the temperature at the liquid-liquid interface is almost constant. More importantly, the liquid-liquid interfacial tension and its variation with temperature are very small in comparison with those at the liquid-gas interface (see table 1).
In order to estimate the influence of the thin lower film on the upper layer, the viscous shear stress at the liquid-liquid interface in the case of two conjugated liquid layers $(h_{01}=\SI{5}{\mum},h_{02}=\SI{510}{\mum})$ has been compared with the shear stress at the wall of the corresponding single upper layer ($h_{02}=\SI{510}{\mum}$), in which a no-slip condition is applied at the bottom. Under the assumption of Couette-type of flow profiles in each layer and constant thermocapillary stress acting at the liquid-gas interface, one finds that the viscous stress at the liquid-liquid interface is reduced only about $10\:\%$ compared to the wall stress in the single layer system (i.e. only the upper layer). In spite of this small reduction in shear stress, the critical Marangoni number of the BM-instability in the upper liquid layer has been found experimentally for the multi-layer system to be approximately $61$ (see figure \ref{fig:Ons_WL}(a)). This value is close to the one of the critical Marangoni number if a free-slip (stress-free) condition is applied at the bottom of the upper layer \citep{BOECK97}. Firstly, this surprising discrepancy is partially due to an assumption made in the simple estimate described above. Postulating a one-way coupling, it was assumed that the thermocapillary stress along the liquid-gas interface remains the same for both cases. By contrast, as it is shown in appendix \ref{appC}, the (horizontal) temperature gradient (and therefore also the thermocapillary stress) at the liquid-gas interface at the onset of the instability in the upper liquid layer changes nonlinearly with the thickness ratio $\bar{h}=h_{02}/h_{01}$. Secondly, the shear stress at the liquid-liquid interface at the onset of the instability is a nonlinear function of the critical Marangoni number, i.e. it is not a good characteristic identifier of the instability in the upper liquid layer. By contrast to the simple estimate and as further supported by the numerical results discussed in appendix \ref{appC}, the experimental findings presented here compare well with the theoretical results obtained for a free-slip condition applied at the bottom of the upper liquid layer. A linear stability analysis of a liquid layer subjected to thermocapillary stresses at the free surface, where the no-slip boundary condition at the wall was replaced by the condition of vanishing stress, was conducted by \citet{BOECK97}. They obtained a neutral stability condition given by
\begin{equation}
\Ma(q,\Biot)=\frac{8q\sinh(q)^2(q\cosh(q)+\Biot\sinh(q))}{\cosh(q)^3+q\sinh(q)-(2q^2+1)\cosh(q)},
\label{eq:fs}
\end{equation}
which to date has not been experimentally verified. The expression indicates that for $\Biot=0$ the instability occurs at $\Ma_c\approx57.59$. Hence, such systems exhibit a lower instability threshold compared to those with a no-slip boundary condition applied at the bottom wall. This behavior can be attributed to the stabilizing effect of wall friction. In the experiments the Biot number is defined by $\Biot=(\kappa_g h_{02})/(\kappa_2 h_g)$, where the subscript $g$ indicates the gas layer. For relatively small values of $h_{02}$ (small $\bar{h}$ at constant $h_{01}$), the experimental findings for the instability threshold of the coupled configuration (shown in figure \ref{fig:Ons_WL}(a)) match with the predictions of equation (\ref{eq:fs}). As shown in appendix \ref{appC}, small $\bar{h}$ implies vanishing stresses at the liquid-liquid interface so that the system with conjugated layers is the first experimentally feasible approach to verify the validity of equation (\ref{eq:fs}). With increasing thickness of the upper liquid layer, the experimental results slowly deviate from the predictions obtained from the free-slip condition. This behavior can be attributed to the vanishing effect of the lubricating layer with increasing $h_{02}$ (larger $\bar{h}$), which asymptotically approaches the no-slip case (see appendix \ref{appC}). 

The reduced friction in the upper liquid layer within the stacked configuration also decreases the dominant wavelength compared to the case of an isolated single film. This is shown in figure \ref{fig:Ons_WL}(b). The average reduction is found to be about $5\% -10 \%$.

\begin{figure}
  \centerline{\includegraphics{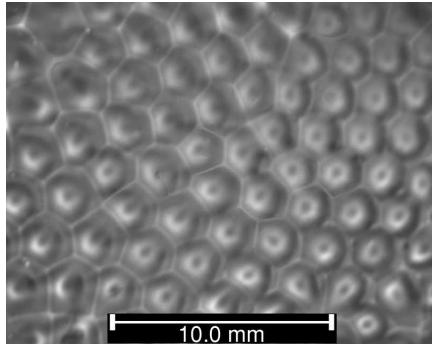}}
  \caption{Pattern formation at the liquid-gas interface of a single layer of FC-70 on a pre-structured substrate. Here, $h_{02}=880\:\mu\textrm{m}$ and $\Ma=1000$. The structures on the lower substrate were produced in a previous step with the multi-layer system, where the lower oil film was replaced by a UV-curable polymer with $h_{01}=5\:\mu\textrm{m}$. The polymer was solidified after the liquid-liquid interface was deformed by the BM-convection in the upper FC-70 layer. In this pre-processing step,the Marangoni number of the upper layer was set to $\Ma=450$. In the figure the dark spots are probably caused by light refraction at the slightly deformed liquid-gas interface. The centers of the convection cells are exactly located at the peaks of the structure of the lower substrate.}
\label{Polymer}
\end{figure}

While the convection rolls in the upper liquid layer shear and deform the liquid-liquid interface, the resulting deformations, albeit small compared to $h_{02}$, in turn significantly affect the temperature distribution at the liquid-gas interface. This occurs because the temperature differences along a convection cell at the liquid-gas interface are within millikelvins only \citep{Dutton10} and are sensitive to the deforming liquid-liquid interface. To demonstrate this, the regular multi-layer setup was first used to induce deformations of the liquid-liquid interface, similar to what is shown in figures \ref{fig:coupl}(b) and \ref{fig:WL}. However, this time a liquid polymer (SEMICOSIL, Wacker) curable by ultra-violet (UV) light was employed instead of silicone oil. After the liquid-liquid interface had been deformed by the shear stresses generated by the BM-convection in the upper layer, UV-light was shone from the top through the viewing window (see figure \ref{fig:ExSet}(a)) to solidify the lower film. The upper layer (FC-70) is not affected by the UV-light. Subsequently, the setup was first brought back to isothermal conditions and, after waiting for several hours, it was subjected to a vertical temperature gradient again. The result is shown in figure \ref{Polymer}. One observes that the convection pattern emerging in the FC-70 layer exactly follows the pattern of the solidified polymer film underneath. The center of each convection cell is in perfect alignment with the peaks of the lower structure. Note that exposing the upper liquid layer to the same temperature gradient without the lower solidified polymer structure being present would not lead to the same arrangement of the convection cells at the liquid-gas interface: The layer thickness and the temperature gradient across it only determine the dominant wavelength of the pattern, but not its exact structure. The accurate pattern alignment indicates that the elevations of the solidified polymer film, although small compared to $h_{02}$, are nevertheless sufficient to affect the structure of the convection pattern. The observed pattern alignment can have two origins: firstly, an initial convection pattern may be forced by hydrodynamic stresses to align with the structure of the solidified polymer film. Secondly, the deformed polymer film may alter the temperature distribution at the liquid-gas interface before that convection sets in. This modulation of the interfacial temperature would cause corresponding modifications of the thermocapillary stress, so that the convection patterns display the same structure as the polymer film. Rough estimates based on a convection-free system suggest that the deformation of the polymer film leads to temperature variations at the liquid-gas interface in the order of $20-100$ millikelvins. This is well within the regime of temperature variations observed in conventional BM-convection in a single liquid layer \citep{Dutton10}. No matter which of the presumed mechanisms dominates, given that the polymer film is a solid after UV-treatment, the complete alignment between the convection cells and the polymer film structure must indeed be caused by the deformations of the polymer. In the same manner, the lower film in the conjugated configuration influences the temperature distribution at the liquid-gas interface and the hydrodynamic field of the upper liquid layer even in an unsolidified state. This enhances the robustness of the pattern against external disturbances, which is amplified with increasing viscosity of the lower film: While a higher viscosity demands a longer time span for the liquid-liquid interface to be deformed by the shear stress induced by the convection cells in the upper liquid layer, the liquid-liquid interface will remain longer in a deformed state after the shear stress exerted from the upper liquid layer is removed. Hence, especially for higher viscosities of the lower film, the deformations at the liquid-liquid interface preserve the information of the pattern in the upper liquid layer. To the best of our knowledge, such a pattern-preserving effect of a lubricating film has never been discussed before.

\section{Mathematical formulation}\label{sec:Math}
\begin{figure}
  \centerline{\includegraphics{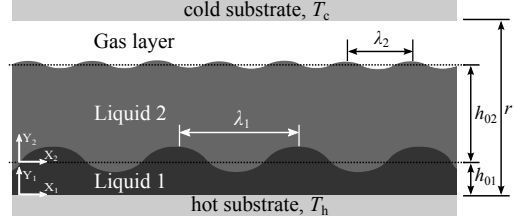}}
  \caption{The configuration of the conjugated liquid system as used in the numerical analysis}
\label{fig:3L}
\end{figure}

In this section the behavior of the coupled system as used in the experiments is modeled and analyzed. Firstly, the two-dimensional governing equations for a liquid layer in contact with a thinner film of a different liquid and subjected to a transverse temperature gradient are specified. Subsequently, the evolution equation for the thin film confined between the solid wall and the thicker liquid layer is derived. The configuration of the stacked liquid layers is schematically shown in figure \ref{fig:3L}. In the lateral direction the system is assumed to be unbounded and the evaporation at the liquid-gas interface is neglected. This is justified since FC-70 has a low vapor pressure at ambient conditions, and the highest temperature does not exceed $30\:^\circ\textrm{C}$. For the low flow velocities and shear rates present in the system under discussion, both liquids are incompressible and have a Newtonian behavior. The system is restricted from the top and the bottom by rigid walls, and the total gap between the two substrates is constant and given by $r$. Furthermore, it is assumed that the two liquid films are immiscible, and the sum of their thicknesses is constant and given by $d_0=h_{01}+h_{02}$.  The lower liquid film is heated from below, while the upper free surface of the thicker liquid film, adjacent to a gas layer of thickness $r-d_0$, is cooled from above. The transverse temperature difference across the three layers of the conjugated system is constant and given by $\Delta T=T_h-T_c$. For $\Delta T>\Delta T_C$ the upper liquid layer undergoes a SW-BM instability, which has a characteristic pattern wavelength $\lambda_2$. Here the upper liquid layer is sufficiently thin so that buoyancy effects and related RB-convection are negligibly small. The dimensionless variables in space $(X_2,Y_2)$, time $\tau$, velocity $(U_2,V_2)$ and temperature $\bar{T}_2$ are scaled by $h_{02}$, $h_{02}^2/\alpha_2$, $\alpha_2/h_{02}$ and $\Delta T_{2}$, respectively. The choice of the characteristic time of thermal diffusion as the dominating time scale is common for problems where weak convection is thermally induced. With this the governing conservation equations for mass, momentum and energy of the upper liquid layer can be brought in the following form:
\begin{subeqnarray}
\frac{\partial\omega}{\partial\tau}+\frac{\partial\Psi}{\partial Y_2}\frac{\partial\omega}{\partial X_2}-\frac{\partial\Psi}{\partial X_2}\frac{\partial\omega}{\partial Y_2} &=& \Pran\left(\frac{\partial^2\omega}{\partial X_2^2}+\frac{\partial^2\omega}{\partial Y_2^2}\right),\\
\frac{\partial\theta}{\partial\tau}+\frac{\partial\Psi}{\partial Y_2}\frac{\partial\theta}{\partial X_2}-\frac{\partial\Psi}{\partial X_2}\frac{\partial\theta}{\partial Y_2}+\frac{\partial\Psi}{\partial X_2} &=& \frac{\partial^2\theta}{\partial X_2^2}+\frac{\partial^2\theta}{\partial Y_2^2},\\
\omega &=& -\frac{\partial^2\Psi}{\partial X_2^2}-\frac{\partial^2\Psi}{\partial Y_2^2}.
\label{SW_equation}
\end{subeqnarray}
Here $\Pran=\nu_2/\alpha_2$ expresses the Prandtl number, while $\Psi$ denotes the stream function defined by $U_2=\partial\Psi/\partial Y_2$ and $V_2=-\partial\Psi/\partial X_2$. In (\ref{SW_equation}a), note that the choice of the characteristic thermal diffusion time as dominating time scale implies the presence of $\Pran$ instead of the Reynolds number (see for instance \citet{Thess95}). The non-vanishing $Z$-component of the vorticity directed outward of the plane is denoted by $\omega$. 
It is assumed that the temperature distribution in the upper liquid layer obeys
\begin{equation}
\bar{T}_2=\frac{T_{l-l}^{cond}}{\Delta T_{2}}-Y_2+\theta,
\end{equation}
where $T_{l-l}^{cond}$ is the temperature of the liquid-liquid interface in the conductive state and $\theta$ denotes the deviation of the temperature from that state. The conductive state refers to the condition where both layers are stable and heat is transfered only through thermal conduction. The boundary conditions at the liquid-gas interface read
\begin{subeqnarray}
\frac{\partial^2\Psi}{\partial Y_2^2}&=&-\Ma\frac{\partial\theta}{\partial X_2},\\
\frac{\partial\theta}{\partial Y_2}&=&-\Biot\:\theta,\\
\frac{\partial\Psi}{\partial X_2}&=&0.
\label{BCLG}
\end{subeqnarray}
The first is the tangential stress balance and the second is the continuity of heat flux. Justified by the large value of the Galileo number (see section \ref{exp_result} and table \ref{tab:DN}, $\Ga\equiv\Ma/\Dy$), it is assumed that the liquid-gas interface is non-deformable. Consequently, the $Y$-component of the velocity at the interface vanishes, which is expressed by the last condition.
\begin{table}
  \begin{center}
\def~{\hphantom{0}}
  \begin{tabular}{llll}
      Dimensionless number   & Symbol&Definition& Typical values \\[3pt]
       Marangoni (l-g)       & \Ma    &$-\gamma_{T,2} h_{02} \Delta T_{2}  /(\mu_2\alpha_2)$&0 $-$ 450\\
       Marangoni (l-l)  	   & $\Ma_1$  &$-\gamma_{T,1} h_{01}\Delta T_{1}/(\mu_1\alpha_1)$&0 $-$ 0.05\\
       Inverse dynamic Bond (l-g)& $\Dy$  &$-\gamma_{T,2} \Delta T_{2}/(\rho_2 g h_{02}^2)$&0 $-$ 0.05\\
       Inverse dynamic Bond (l-l)& $\Dy_1$  &$-\gamma_{T,1} \Delta T_{1}/(\rho_1 g h_{01}^2)$&0 $-$ 1.5\\
       Rayleigh							 &\Ra& $\beta g h_{02}^3\Delta T_2/(\nu_2\alpha_2)$&0 $-$ 45\\
       Biot                  & \Biot  &$\kappa_g h_{02}/(\kappa_2 h_g)$&0 $-$ 1 \\
       Bond                  & $\Bo$&$(\rho_1-\rho_2) g h_{01}^2/\gamma_{1}$&-4.6$\times 10^{-5}$ \\
       Prandtl               & \Pran  &$\nu_2/\alpha_2$&365\\
       Hamaker               & $A $&$A_{sfo}/(6\pi\gamma_{1} h_{01}^2)$&-1.13$\times 10^{-8}$\\
       Capillary             & $\Ca$&$\mu_1\alpha_1/(\gamma_{1}h_{01})$&8.3$\times 10^{-3}$\\
       
  \end{tabular}
  \caption{Typical values of dimensionless numbers}
  \label{tab:DN}
  \end{center}
\end{table}

As stated before, the convection cells emerging from the BM-instability shear and deform the liquid-liquid interface with a characteristic wavelength of $\lambda_1\approx\lambda_2$. Since the lower film is very thin, this wavelength is much larger than $h_{01}$. Hence any deformations emerging at the liquid-liquid interface have a LW-character. Numerous authors have derived and analyzed the LW-evolution equation of a thin liquid layer deposited on a solid substrate and subjected to a transverse temperature gradient. However, the vast majority of these studies considers the case where the film is adjacent to a gas of negligible viscosity. In the case of the conjugated liquid layers the shear stress as well as the pressure exerted by the upper liquid layer on the liquid-liquid interface have to be taken into account. Here, only the final evolution equation will be presented. For a detailed description of the LW-evolution equation derived under the lubrication approximation, readers are referred to \cite{Oron97,Pototsky05} and \citet{Merkt2005}. Accordingly, in dimensionless form, the lubrication approximation of the thin liquid film bounded by a thicker film reads
\begin{equation}
\frac{\partial H}{\partial\tau}+\frac{\partial}{\partial X_1}\left(\frac{H^2}{2}\left(\frac{\partial U_1}{\partial Y_1}\right)_{l-l}-\frac{\bar{\mu}H^3}{3\bar{h}^2}\left(\frac{\partial P_1}{\partial X_1}\right)_{l-l}\right)=0.
\label{evolution}
\end{equation}
Here the scaling of the vertical coordinate and the local film thickness $h$ are based on the thickness of the lower liquid film, i.e. $Y_1=y_1/h_{01}$ and $H=h/h_{01}$, while the lateral coordinate, time, velocity and pressure are scaled using the characteristic values of the upper liquid layer. The dimensionless pressure in the lower liquid film is scaled according to $P_1=h_{02}^2(p_1+\phi_1)/(\mu_2\alpha_2)$, where $\mu_2$ is the dynamic viscosity of the upper liquid layer, $p_1$ is the (dimensional) liquid pressure and $\phi_1$ represents the sum of the hydrostatic and the disjoining pressure in the lower film. The ratios of the initial film heights and the dynamic viscosities of the upper liquid layer and the lower one are denoted by $\bar{h}=h_{02}/h_{01}$ and $\bar{\mu}=\mu_2/\mu_1$, respectively. Note that the lubrication approximation implies that the vertical pressure gradient in the lower film vanishes, i.e. the pressure in the interior of the lower film at any horizontal position is the same as at the corresponding horizontal location at the liquid-liquid interface, which is (mainly) balanced by the capillary pressure of that interface. This is the reason why a shear stress-induced deformation of the lower film can be observed: at steady state, the internal recirculation in a liquid layer requires a horizontal pressure gradient in the interior of the layer; for thin films governed by the lubrication approximation, this horizontal pressure gradient is also present at the liquid-liquid interface and causes a gradient in capillary pressure and corresponding interfacial deformations. Such arguments cannot be employed for thicker lower films not governed by the lubrication approximation. However, for thicker lower films, the liquid-liquid interface may deform due to the LW-RT instability alone, see the earlier discussion of this topic. The dimensionless pressure gradient at the liquid-liquid interface is given by  
\begin{equation}
\left(\frac{\partial P_1}{\partial X_1}\right)_{l-l}=\frac{-\bar{h}^2}{\bar{\mu}\bar{\alpha}\Ca}\left(\frac{1}{\bar{h}^2}\frac{\partial^3 H}{\partial X_1^3}-\Bo \frac{\partial H}{\partial X_1}+\frac{3A}{H^4}\frac{\partial H}{\partial X_1}\right)+\left(\frac{\partial P_2}{\partial X_1}\right)_{l-l}.
\label{Dpressure}
\end{equation}
Here $P_2$ is a dimensionless reference pressure, which - in the present case - is the pressure in the upper liquid layer and scaled according to $P_2=h_{02}^2(p_2+\phi_2)/(\mu_2\alpha_2)$. The ratio of the thermal diffusivities of the upper liquid layer and the lower one is denoted by $\bar{\alpha}=\alpha_2/\alpha_1$. The Capillary number reads $\Ca=\mu_1\alpha_1/(\gamma_{1}h_{01})$ and the Bond number is given by $\Bo=(\rho_1-\rho_2) g h_{01}^2/\gamma_{1}$, where $\gamma_{1}$ denotes the interfacial tension between the two liquids. Moreover, the dimensionless Hamaker parameter reads $A=A_{sfo}/(6\pi\gamma_{1} h_{01}^2)$, where $A_{sfo}$ denotes the Hamaker constant specifying the (molecular) interaction strength between the silicon wafer-silicone oil (s-o) and the silicone oil-FC-70 (f-o) interfaces. The first term enclosed by the first parentheses on the right-hand side of equation (\ref{Dpressure}) is the gradient in capillary pressure, while the second term expresses the gradient in hydrostatic pressure of the lower film. The influence of the disjoining pressure arising from van der Waals interactions is captured by the third term. Since the Prandtl number of the upper layer is large (see table \ref{tab:DN}), the left-hand side of equation (\ref{SW_equation}$a$) can be neglected, which simplifies these expressions into the Stokes equations. In this limit, the pressure gradient can be approximated by 
\begin{subeqnarray}
\frac{\partial P_2}{\partial X_2}&=&-\frac{\partial\omega}{\partial Y_2},\\
\frac{\partial P_2}{\partial Y_2}&=&\frac{\partial\omega}{\partial X_2}.
\label{Pr_2}
\end{subeqnarray}

According to the results obtained from the experiments, the silicone oil always wets the whole area of the silicon wafer, even at the rims of the hexagonal cells, where the thickness of the silicone oil may become very small. The disjoining pressure prevents dewetting of the silicone oil from the substrate and stabilizes the liquid film. To calculate $A_{sfo}$, only the long-range interaction forces have been considered, since the contribution of short-range interactions have to be included only for ultrathin films of thicknesses below $\SI{10}{nm}$ \citep{colinet2010pattern}. Since the thickness of the upper liquid layer is large compared to the one of the lower film, the contribution of the liquid-gas interface to the long-range interactions is neglected. The Hamaker constant is calculated based on the Lifshitz theory, which yields $A_{sfo}=-2.92\times 10^{-20}\:\textrm{J}$ \citep{israelachvili2011intermolecular}. Some further details of the calculation are given in appendix \ref{appB}. 

Along the solid substrate, the temperature is assumed to be constant and equal to $T_h$, while no-slip and no-penetration are enforced by setting both velocity components to zero. A symmetry boundary condition is applied at the lateral domain boundaries. The boundary conditions at the liquid-liquid interface couple the governing equations of the upper liquid layer, given by (\ref{SW_equation}), with the evolution equation of the lower film, (\ref{evolution}). On the one hand, this is the continuity of the $X$-component of the velocity, i.e. $U_1(Y_1)|_{Y_1=1}=U_2(Y_2)|_{Y_2=0}$, where, according to the lubrication approximation, the lateral velocity in the lower liquid film is expressed by 
\begin{equation}
U_1=\frac{\bar{\mu}}{\bar{h}^2}\left(\frac{Y_1^2}{2}-HY_1\right)\frac{\partial P_1}{\partial X_1}+\left(\frac{\partial U_1}{\partial Y_1}\right)_{l-l}Y_1.
\label{xvelocity}
\end{equation}
On the other hand, this is a jump of the tangential stress across the liquid-liquid interface given by
\begin{equation}
\left(\frac{\partial U_1}{\partial Y_1}\right)_{l-l}=-\frac{\overline{\Delta T}\mbox{\textit{Ma}}_1}{\bar{\alpha}}\left(\frac{\partial\theta}{\partial X_1}-\frac{\bar{\kappa} D_0}{(D_0-H)^2}\frac{\partial H}{\partial X_1}\right)+\frac{\bar{\mu}}{\bar{h}}\left(\frac{\partial U_2}{\partial Y_2}\right)_{l-l}.
\label{stressjump}
\end{equation}
Here, $D_0=d_0/h_{01}$ and $\bar{\kappa}=\kappa_2/\kappa_1$ denotes the ratio of the thermal conductivities of the films. The ratio of the temperature differences across the upper layer and the one across the lower film is expressed by $\overline{\Delta T}=\Delta T_2/\Delta T_1$. The liquid-liquid Marangoni number is expressed by $\mbox{\textit{Ma}}_1=-\gamma_{T,1} h_{01}\Delta T_{1}/(\alpha_1\mu_1)$, while $\gamma_{T,1}$ denotes the corresponding Marangoni coefficient. The first term on the right-hand side of equation (\ref{stressjump}) proportional to $\Ma_1$ represents the thermocapillary stress at the liquid-liquid interface, where the temperature modulation is caused by two contributions: The first term in the parentheses expresses the change of temperature at the liquid-liquid interface caused by the convection in the upper layer, while the second term represents the change of the interfacial temperature caused by the deformations of the liquid-liquid interface. Since the deformation at the liquid-liquid interface is much smaller than the thickness of the upper liquid layer, the effect of the deformation on the flow field of the upper layer is neglected. Therefore, the $Y$-component of the velocity at the liquid-liquid interface in the upper liquid layer is set to zero and the flow domain remains rectangular. Note that this assumption is particularly valid if the system is in the vicinity of marginal stability. In the following, the numerical simulations will be mainly concerned with this regime.

The temperature distribution in the lower thin film can be approximated by the solution of a one-dimensional heat conduction problem across the film (see for example  \citet{Dietzel2010} and references therein). Hence, next to the continuity of the interfacial temperature, the continuity of the heat flux at the liquid-liquid interface can be enforced by
\begin{equation}
\frac{T_h-T_{l-l}^{cond}}{\Delta T_{2}}-\theta=-\frac{\bar{\kappa}}{\bar{h}}\left(\frac{\partial\theta}{\partial Y_2}-1\right).
\label{thermalBC}
\end{equation}
 
The set of equations (\ref{SW_equation}) and (\ref{evolution}), along with the corresponding boundary conditions, have been solved employing a Galerkin finite element method with Lagrangian shape functions (quadratic element order). The dimensionless simulation domain is a rectangle with the size of $20$x$1$, which is scaled using the initial thickness of the upper liquid layer. The governing equations (\ref{SW_equation}) are solved inside the domain, whereas the evolution equation (\ref{evolution}) is solved along the lower domain boundary. The equations are implemented utilizing the Partial Differential Equation (PDE) mode of COMSOL 4.4 \citep{comsol}. The initial thickness of the lower film is $h_{01}=$ \SI{5}{\mum}, while the initial thickness of the upper liquid layer ranges from $h_{02}=$ \SI{500}{\mum} to \SI{1000}{\mum}. To ensure that the solution reaches a steady-state, transient simulations were carried out until no further change of the dependent variables was observed. For all simulation results shown, this was accomplished by computing the time evolution up to a dimensionless time $\tau=250$. The simulations were performed on a Dell Precision T7500 workstation using CentOS 5.11 as operating system. A mesh convergence study was conducted by refining the mesh in four steps from $100$x$10$ to $400$x$40$, while evaluating the variations of the flow pattern wavelength as well as of the temperature at the center of the convection cells for each refinement step. After the second mesh refinement the relative variation of the parameters reached $0.5\%$, which indicates numerical convergence with respect to the mesh size beyond $200$x$20$ grid points. This resolution was used in all simulations presented herein. A convergence study with respect to the time step was conducted as well. Since the COMSOL solver employs a variable time-stepping scheme according to a specified relative tolerance, such a convergence study can be conducted by either successively reducing the relative tolerance allowed within the iterations for one time step or by successively reducing the maximum time step the solver may select. The second option was chosen herein, where it was ensured that the tested maximum time steps are within the range of values from which the solver automatically chooses. To this end, the maximum value of the dimensionless time step was enforced to be either $0.1$, $0.2$ or $0.5$. For dimensionless times $\tau=9$ and $\tau=250$ the simulation results showed that for maximum time steps smaller than $0.5$ the change of the temperature at the center of the convection cells as well as the modification of the pattern wavelength was negligibly small. Thus, in all of the simulations performed, the maximum dimensionless time step was set to $0.2$. For initialization, the lower film  was perturbed by a random fluctuation of the initial film thickness with an amplitude of $5\times10^{-6}$. 

\begin{figure}
  \centerline{\includegraphics{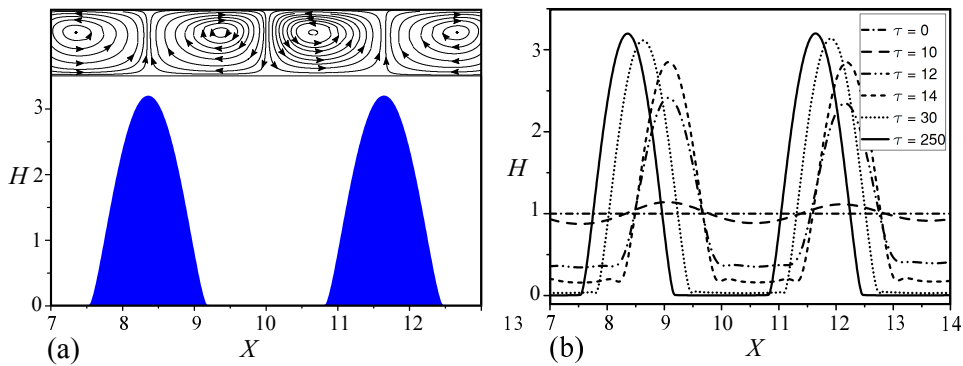}}
  \caption{(a) Shear-driven LW-deformation of the lower film at $\tau=250$. In the upper liquid layer the $Y$-dimension is rescaled for illustrative purposes to fit into the graph. The streamlines in the upper liquid layer and the deformation of the lower liquid layer are depicted for $\mbox{\textit{Ma}}=83$, $h_{02}=600\:\mu\textrm{m}$ and $h_{01}=5\:\mu\textrm{m}$ (b) Time evolution of the liquid-liquid interface for the same parameters as used in (a).}
\label{fig:Deform}
\end{figure}

A typical simulation result at steady-state ($\tau=250$) is shown in figure \ref{fig:Deform}(a). The peaks of the deformations developing at the liquid-liquid interface are located exactly at the center of the convection cells emerging in the upper liquid layer, so that the patterns are aligned. While the film thickness of the regions at the cell centers increases, the lower liquid layer is almost depleted near the edges (see figure \ref{fig:Deform}(b)). At these locations dewetting is prevented by the disjoining pressure. As apparent from the time snapshots of the film evolution shown in figure \ref{fig:Deform}(b), the horizontal location of the deformation pattern shifts with time. This specific behavior is caused by the evolution of the convection cells emerging in the upper liquid layer, which is transient at the beginning of the process. In appendix \ref{appA}, the model presented herein is used to show that the LW-RT instability (related to the density inversion of the liquid system) as well as the LW-BM instability evolve much slower at the liquid-liquid interface than the shear-driven deformation due to the SW-BM convection cells. This is a direct consequence of the smallness of $h_{01}$ used in this study. Along this line, it is also demonstrated that LW-RT and LW-BM induced deformations remain negligibly small throughout the time frame of interest herein.
\begin{figure}
  \centerline{\includegraphics{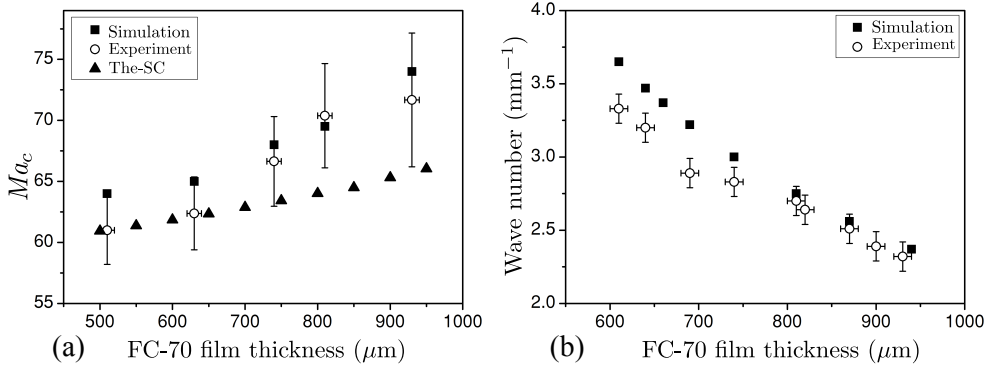}}
  \caption{Comparison of the experimental findings with the results from the mathematical model. (a) Onset of instability (marginal stability). The results of the analytical model for the free-slip condition at the bottom of the upper liquid layer (The-SC) are displayed as well. (b) Dominating pattern wave number at $\Delta T_{2}=3\:\textrm{K}$.} 
\label{fig:Comp}
\end{figure}

In order to put these numerical findings into perspective, the results obtained for the onset of the instability in the upper liquid layer have been compared with the experimental results. This is shown in figure \ref{fig:Comp}(a). The experimental findings indicate that the critical Marangoni number $\Ma_c$ increases as the thickness of the upper liquid layer $h_{02}$ increases (equivalent to increasing $\bar{h}$, since $h_{01}=\SI{5}{\mum}$). Within the experimental uncertainties, this trend is captured correctly by the numerical simulation, whereas the stress-free model of \citet{BOECK97} does not predict this trend. In our study, the maximum experimental uncertainty of $\Ma_c$ (obtained for $h_{02}=930$) is $\pm5.4$, which is well within the typical range of experimental uncertainty observed for such experiments \citep[see][]{Schatz95}. Note that the experimental results, including the error bars, are identical in both figures \ref{fig:Ons_WL}(a) and \ref{fig:Comp}(a), while the range of critical Marangoni numbers along the ordinate of figure \ref{fig:Comp}(a) is smaller. An analogous comparison between experimental and numerical results was conducted for the dominating wave number of the emerging pattern, which is shown in figure \ref{fig:Comp}(b). Also this comparison indicates satisfactory agreement between numerical and experimental results. This supports the presumed mechanism, namely shear-driven deformation of the lower layer coupled to the SW-BM convection in the upper layer.

\section{Conclusions}
In this article, the coupling between two conjugated, immiscible liquid layers of vastly different thicknesses, exposed to a transverse temperature gradient, was investigated. The liquid layers are situated on a hot solid substrate and separated from the colder upper solid substrate by an air layer. The upper thicker liquid layer is seen to develop the short-wavelength (SW) mode (convective mode) of the B\'enard-Marangoni (BM) instability, whereas the growth rates of the long-wavelength (LW, deformational) BM and Rayleigh-Taylor (RT) instabilities of the much thinner film at the bottom, exhibiting a lower density than the upper one, are very small. Thus, the corresponding deformations of the liquid-liquid interface are insignificant within the time frame of interest. The viscous coupling between the two layers mainly follows a master-slave configuration so that the occurrence of the BM-instability in the upper liquid layer triggers a LW-deformational mode in the lower one. Since the BM-instability in the upper layer is accompanied by convection cells resembling an l-type instability, the corresponding shear stresses shift liquid mass of the lower film towards the (horizontal) center of each cell. Hence, the pattern developing at the liquid-liquid interface possesses the same wavelength as the pattern arising within the upper layer. The coupling between the liquid layers leads to a number of differences in the behavior of the system compared to a single film setup. Firstly, the presence of the lower film shifts the marginal stability point of the upper liquid layer to smaller values of the critical Marangoni number $\Ma_c$. For small values of the thickness ratio (upper to lower) $\bar{h}$, $\Ma_c$ is found to be almost as low as theoretically predicted for a free-slip boundary condition applied at the bottom of the upper layer. This is surprising, since a simple model predicts that for equally small $\bar{h}$, the presence of the lubricating film reduces the shear stress at the bottom of the upper layer by only about $10\%$. As $\bar{h}$ increases, $\Ma_c$ approaches its classical value for the case when a no-slip boundary condition is applied at the bottom of the upper layer (equivalent to a single liquid layer). Secondly, the pattern in the coupled system was analyzed using Fourier transformation and compared to the single film setup. It was found that for a constant temperature difference the wavelength of the pattern in the conjugated system is $5\% - 10\%$ smaller than the wavelength in the corresponding single film. Thirdly, it was experimentally proven that the emerging deformations, albeit small, and not the reduced viscous stresses at the liquid-liquid interface increase the robustness of the hexagonal convection cells in the upper liquid layer against external disturbances, which indicates that the coupling between the layers is indeed mutual. These three effects of a thin, lubricating film placed underneath a thicker liquid layer subjected to a transverse temperature gradient appear to have never been explicitly discussed before. The evolution of the pattern in the conjugated system was numerically analyzed. A good overall agreement with the experimental results was found, and the dominating mechanism - with the upper liquid layer driving the deformation of the lower one - was confirmed. The presented system demonstrates that circular convection patterns developing in the upper layer can be used to actively shape a second liquid film by means of LW-deformations. Given that the liquid-liquid interfacial tension and the corresponding capillary stabilization are small, the shaping of the lower film exactly follows the horizontal scale of the convection pattern in the upper layer. This can not only be used as a visualization method for the convection cells (note that the deformations can be seen without any optical or other forms of assistive instruments, whereas the visualization of the flow itself usually requires the addition of tracer particles) but might be also applicable as a novel method for creating highly regular, microstructured surfaces. The presented study is one of the first where a periodic pattern is generated in a hydrodynamic sub-system, without this sub-system having to be intrinsically unstable. Furthermore, for sufficiently small $\bar{h}$, it serves as the first feasible experimental system able to validate the reduction of $\Ma_c$ theoretically predicted for vanishing stress at the bottom of the thicker layer.

\acknowledgments
Funding by the German Research Foundation (DFG), Grant No. DI 1689/1-1, is gratefully acknowledged. The authors thank Arthur Fast for his help with the image analysis.

\appendix
\section{}
\label{appA}
\begin{figure}
  \centerline{\includegraphics{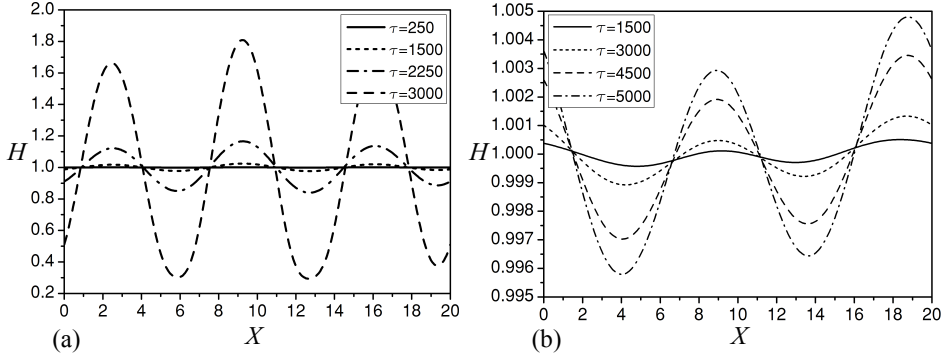}}
  \caption{(a) Time evolution of the liquid-liquid interface in the absence of shear stress exerted by the upper layer for $\Bo=0$, $\Ca=8.3\times10^{-3}$ and $\Dy_1=1.43$ ($\Ma_1=0.0067$) (b) Time evolution of the liquid-liquid interface in the absence of the transverse temperature gradient for $\Bo=-4.6\times10^{-5}$, $\Ca=8.3\times10^{-3}$ and $\Dy_1=0$ ($\Ma_1=0$).}
\label{fig:RT}
\end{figure}
As the density of the lower film is smaller than the one of the upper layer (see table \ref{tab:pp}), the system under study is density-inverted. Therefore, the RT instability may contribute to the evolution of the lower film. Neglecting for this analysis the convection cells in the upper liquid layer, the liquid-liquid interface undergoes the LW-BM instability along with the LW-RT instability. In this case, the time evolution of the lower liquid film thickness is given by
\begin{equation}   
\frac{\partial H}{\partial\tau}+\frac{\partial}{\partial X_1}\left\{\frac{H^2}{2}\left(\frac{\overline{\Delta T}\bar{\kappa}D_0\Ma_1}{\bar{\alpha}(D_0-H)^2}\frac{\partial H}{\partial X_1}\right)+\frac{H^3}{3\bar{\alpha}\Ca}\left(\frac{1}{\bar{h}^2}\frac{\partial^3 H}{\partial X_1^3}-\Bo \frac{\partial H}{\partial X_1}\right)\right\}=0.
\label{evolution_RT}
\end{equation}
The initial film thickness is $h_{01}=5\:\mu\textrm{m}$. Here, we are primarily interested in the time scale of the instability within the linear regime, where deformations are small. Hence, for simplicity, disjoining pressure effects have been omitted from equation (\ref{evolution_RT}). This expression has been solved with the same methodology as outlined in section 3. The film is perturbed initially by a random fluctuation with an amplitude (relative variation of $H$) of $10^{-2}$. The time evolution of the liquid-liquid interface undergoing either the LW-BM or the LW-RT instability is shown in figure \ref{fig:RT}(a) and (b), respectively. In each figure, the respective instability is considered isolated from the other, i.e. in (a), $\Dy_1=1.43$, while gravity is excluded. By contrast, in (b), $\Bo=-4.6\times10^{-5}$, while $\Dy_1=0$. The growth rate of the deformations evolving under the LW-RT instability is very small in comparison to the one of the LW-BM instability. For instance, this can be observed at $\tau=3000$ (i.e. approximately $9$ hours after the initial state in real time), when the amplitude of the deformations due to the LW-RT instability has grown only by $0.1\%$, while this value is much higher for the case of the LW-BM instability. Thus, the contribution of the LW-RT instability in the evolution of the liquid-liquid interface is much smaller than the one of the LW-BM instability. However, in comparison with the time evolution of the deformation of the liquid-liquid interface in the presence of the convection cells in the upper liquid layer (figure \ref{fig:Deform}(b)), it is apparent that even the deformations caused by the LW-BM instability grow very slowly. Therefore, not only the LW-RT but also the LW-BM instability is insignificant for the deformation of the liquid-liquid interface in the time frame of interest. 

This finding, according to which especially the LW-RT instability present in the system under study exhibits a small growth rate, confirms experimental observation. After waiting for several hours with the temperature gradient being absent, no specific changes at the liquid-liquid interface could be observed in the experiment.

\section{}
\label{appC}
\begin{figure}
  \centerline{\includegraphics{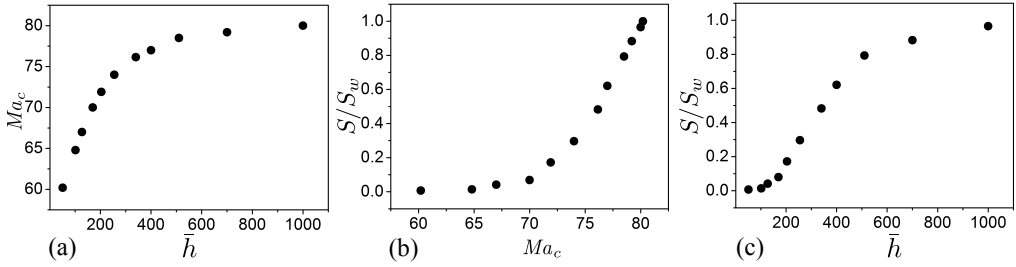}}
  \caption{(a) Numerical results for the critical Marangoni number in the upper liquid layer as a function of $\bar{h}=h_{02}/h_{01}$ (b) Shear stress $S$ at the bottom of the upper liquid layer, determined at the onset of the BM-instability and scaled to its maximum value $S_w$ obtained for the no-slip condition, as a function of the critical Marangoni number in the upper liquid layer. In dimensionless form, the shear stress takes the maximum value of $\chi_{w}=S_w h_{02}^2/(\mu_2\alpha_2)=1.45$, if the no-slip condition is used at the bottom of the upper layer. (c) Ratio $S/S_{w}$ as a function of $\bar{h}$. All numerical results shown in these three plots were obtained for $h_{02}=\SI{510}{\mum}$, $\bar{\mu}=8.39$ and $\Biot=0.108$.}
\label{fig:PS}
\end{figure}

On the one hand, the comparison of the experimental results of $\Ma_c$ with the theoretical work of \citet{BOECK97} indicates that at marginal stability for $\bar{h}= h_{02}/h_{01} < 130$, the shear stress at the bottom of the upper liquid layer practically vanishes in the conjugated configuration (see figure \ref{fig:Comp}(a) with $h_{01}= \SI{5}{\mum}$). On the other hand, as mentioned in the main text, a simple estimate based on Couette-type flow profiles rather suggests that the presence of the lubricating film reduces this stress by only about $10\:\%$ with respect to the wall stress in a corresponding single layer system. Presumably, this mismatch is caused by the underlying assumption used in the simple analysis that the thermocapillary stress at the liquid-gas interface of the upper liquid layer is not affected by the lubricating film. Despite of the numerous works devoted to the BM-instability, no analytical or numerical study has been carried out on the temperature distribution (and thus on the thermocapillary stress) at the liquid-gas interface at the onset of the SW-instability. To confirm the presumed shortcoming of the simple estimate and address the effect of the depth ratio $\bar{h}$, some additional simulations have been performed. Here the thickness of the upper liquid layer remains constant at $h_{02}=\SI{510}{\mum}$, while the thickness of the lower layer varies. All the simulations have been carried out based on the methodology described in section $3$. As shown in figure \ref{fig:PS}(a), the instability threshold of the upper liquid layer, quantified by $\Ma_c$, changes nonlinearly with the depth ratio $\bar{h}$. Since for $\bar{h}$ smaller than approximately $10$ the lubrication approximation in the lower liquid film is not valid anymore, the numerical simulations have been performed for corresponding larger values of $\bar{h}$ ($10\leq\bar{h}\leq1000$). At values of $\bar{h}$ smaller than $\approx 100$ the critical Marangoni number is close to $61$, which corresponds to the case of a free-slip boundary condition applied at the bottom of the upper liquid layer \citep{BOECK97}. It should be noted that in the simulations $\Biot=0.108$, which tends to increase $\Ma_c$ compared to the case of $\Biot=0$. As $\bar{h}$ increases, the instability threshold approaches the value of $\Ma_c=81$, which is the classical value at the point of marginal stability obtained for the case that a no-slip boundary condition is applied at the bottom of the upper layer. Then, as shown in figure \ref{fig:PS}(b), the exerted shear stress at the liquid-liquid interface, denoted by $S$, takes its maximum value $S_w$, whereas at $\Ma_c\approx61$ $S$ is indeed vanishingly small. In this plot it is also interesting to see that the ratio $S/S_w$ varies very little in the range of $60<\Ma_c <70$, i.e. the shear stress at the liquid-liquid interface remains vanishingly small. For larger values of $\Ma_c$ it rapidly increases to $S_w$. As shown in figure \ref{fig:PS}(c), a similar behavior can be observed if $S/S_w$ is plotted as a function of $\bar{h}$: By increasing $\bar{h}$ from small values up to $200$, the dimensionless shear stress at the liquid-liquid interface increases from practically zero to only about $15\%$ of the maximum value. For $200<\bar{h}<500$, the shear stress at the liquid-liquid interface rapidly increases, while for $\bar{h}>500$ the value of $S_w$ is  approached asymptotically.

\section{}
\label{appB}
The Hamaker constant has been calculated based on the Lifshitz theory, which assumes that the main absorption frequencies of all involved media are about $\nu_e=3\times10^{15}\:\textrm{Hz}$, and especially that zero-frequency contributions are negligible \citep{israelachvili2011intermolecular}. An approximation of the non-retarded Hamaker constant of the two macroscopic phases FC-70 (f) and silicon (s) interacting across the silicone oil (o) layer is given by
\begin{equation}
A_{sfo}\approx\frac{3h\nu_e}{8\sqrt{2}}\frac{(n_s^2-n_o^2)(n_f^2-n_o^2)}{\sqrt{n_s^2+n_o^2}\sqrt{n_f^2+n_o^2}\left(\sqrt{n_s^2+n_o^2}+\sqrt{n_f^2+n_o^2}\right)} ,
\end{equation}
where $n_i$ denote the refractive indices and $h=6.626\times10^{-34}\:\textrm{Js}$ represents Planck's constant. Using refractive indices according to $n_s=3.44$, $n_f=1.308$ and $n_o=1.39$, the non-retarded Hamaker constant is computed as $A_{sfo}=-2.92\times 10^{-20}\:\textrm{J}$.

\bibliographystyle{jfm}

\bibliography{ref}
\end{document}